\documentstyle[amssymb,12pt]{article}

\input{tcilatex}

\begin{document}

\title{QUANTUM SUPERPOSITION OF PARAMETRICALLY AMPLIFIED MULTIPHOTON PURE STATES
WHITIN A DECOHERENCE-FREE SCHROEDINGER-CAT STRUCTURE}
\author{F.A. Bovino, F. De Martini and V. Mussi \\
Dipartimento di Fisica Universit\`{a} ''La Sapienza'' di Roma, Italia}
\maketitle

\begin{abstract}
The new process of {\it quantum-injection} into an optical parametric
amplifier operating in {\it entangled} configuration is adopted to {\it %
amplify} into a large dimensionality spin-$%
{\frac12}%
~$Hilbert space the quantum entanglement and superposition properties of the
photon-couples generated by parametric down-conversion. The structure of the
Wigner function and of the field's correlation functions shows a {\it %
decoherence-free}, multiphoton {\it Schroedinger-cat} behaviour of the
emitted field which is largely detectable against the squeezed-vacuum noise.
Furthermore, owing to its entanglement character, the system is found to
exhibit multi-particle quantum nonseparability and Bell-type nonlocality
properties. These relevant quantum features are analyzed for several {\it %
travelling-wave} optical configurations implying different input
quantum-injection schemes. (PACS numbers: 03.65.Bz, 03.67.-a, 42.50.-p,
89.70.+c).
\end{abstract}

\section{ Introduction}

The generation of classically distinguishable quantum states, a major
endeavor of modern physics, has long been the object of extensive
theoretical studies. In recent times important experimental investigation
with atoms has been carried out in this field by various research groups 
\cite{Schroedinger}\cite{caldeira}\cite{monroe}\cite{haroche}\cite{noel}. In
this context it has been proved that the realization of the Schroedinger-cat
program is generally challenged by an extremely rapid decoherence process
due to the stochastic interactions of any freely evolving{\it \ }mesoscopic
system with the environment \cite{daniel}\cite{zurek}. Within the framework
of quantum computation{\it \ }the same process has also been recognized to
represent a major limitation toward the coherent superposition of the qubits
carrying the quantum information \cite{shor}. In the domain of quantum
optics several strategies have been proposed to overcome the problem, e.g.
the back-action evasion \cite{caves} and cavity control by optical feedback 
\cite{milburn}\cite{tombesi}. \ In the present letter we present a new
approach to the problem based on the amplifying / squeezing operation of the
travelling wave optical parametric amplifier (OPA) operating in a novel
entangled configuration and initiated by a dynamical interaction process
here introduced for the first time in the framework of the nonlinear (NL)
parametric amplification: {\it quantum-injection. }In general we may define
the quantum injection process in connection with any amplifying or
scattering system as the one provided by an input field whose $%
P-Representation$ does not exist as a tempered solution \cite{walls}. In our
present case the character of {\it quantum-injection \ }is provided by the
subpoissonian character of a single photon in the Fock state n=1 in a
quantum superposition of polarization, or momentum, states. Sometimes we
refer to this single particle state as the input {\it qubit}. This photon
may belong to a couple generated by spontaneous parametric down-conversion
(SPDC) e.g. in a $\Phi -phase\ tunable$ entangled state of linear
polarization ${\bf \pi ,}$defined in a Hilbert space of dimensionality $%
2\times 2$. The SPDC process has been adopted within recent tests of
violation of Bell'inequalities \cite{mattle}, of quantum state-teleportation 
\cite{boschi} and of all processes generally belonging to the chapter of 
{\it nonlocal entangled interferometry }\cite{q-if}\cite{branca}. The key
idea of the present work relates to the possibility of ``amplifying'' this
quite interesting phenomenology to a higher dimensionality spin-$%
{\frac12}%
$ Hilbert space, i.e. involving a large number of photon couples, by taking
advantage of the {\it unitary }character of the transformation accounting
for the amplification process. We show that this can be realized by a novel
optical device, the quantum-injected{\it ,} {\it entangled OPA} leading to a
new {\it entangled} Schroedinger-cat (S-cat) configuration which may be {\it %
decoherence-free}, in the ideal case. A first account of the present work is
found in Refs.\cite{Fdemartini}. This quite interesting condition, implying
the linear superposition of two multi-particle, i.e. macroscopic{\it \
pure-states} will be investigated for two different optical travelling-wave
configurations: (1) the quantum-injected non-degenerate OPA and (2) the
quantum-injected mode-degenerate OPA. Both schemes will be analyzed
theoretically by two different and complementary approaches. Accordingly,
the paper is organized as follows:

First, the dynamical unitary evolution of the input qubit providing the
quantum injection will be analyzed in the details for the two configurations
in Sections 2 and 3, respectively. Second, the general approach will be
followed by an {\it exact}, closed form evaluation of the Wigner function,
in Sections 4 again for both configurations. The details of all calculations
will be given for the more complex and elaborate configuration (1), for
generality.Rather surprisingly, we shall see that the formal expression of
the Wigner function is found the same for both configurations in spite of
the somewhat different dynamics and of the different signification of the
variables appearing in the resulting expression. This may emphasize the
formal role taken by the common single-photon quantum injection scheme
within the quantum analysis. In any case this allows an interesting unifying
Wigner function analysis showing, for both configurations, the relevant
multi particle quantum superposition properties of the S-cat. Third, a
theory of the {\it first-order} optical correlation functions of the
parametrically generated field, given in Section 5 suggests, for both cases,
a straightforward first-order interferometric method for a direct
single-beam detection of the S-cat. In addition, the detailed theoretical
investigation of the {\it second-order} inter-mode correlation function
leads to a Bell-type multi-photon quantum nonlocal behaviour of the emitted
field. At last, a new multi-particle Bell-inequality experiment will be
considered.

\section{Quantum Injection in the non-degenerate Optical Parametric
Amplifier. Amplification of quantum entanglement.}

Consider the diagram shown in Figure 1, a diagram suggested by an actual
experimental investigation presently being carried out in our Laboratory.
Two equal and equally oriented NL\ crystals, e.g., beta-barium-borate (BBO)
cut for Type II phase matching are excited by two beams derived from a
common UV\ laser beam at a wavelength (wl) $\lambda _{p}=2\pi \left| {\bf k}%
_{p}\right| ^{-1}$. Crystal 1 is the SPDC source of ${\bf \pi -}${\it %
entangled} photon couples emitted, with wl $\lambda =2\lambda _{p}$ over the
modes ${\bf k}_{1\text{,}}$ ${\bf k}_{3}$ determined by two fixed pinholes.

\FRAME{fhFU}{272.4375pt}{210.4375pt}{0pt}{\Qcb{Optical configuration of the
quantum-injected, entangled optical parametric amplifier realizing the
process of multiphoton quantum superposition. The SPDC quantum injector is
provided by a type II $\Phi $-phase tuneble generator of linear polarization
($\protect\pi $)-entangled photon couples. The crystal realizing the OPA
action is cut for type II, noncollinear phase matching and is equal to the
one realizing SPDC. The detection system consists of the of a birefringent
plate $\Psi $, a $\protect\pi $ rotator R($\protect\varphi $), a polarizing
beam-splitter PBS and two cooled photomultipliers.}}{}{Figure }{\special%
{language "Scientific Word";type "GRAPHIC";maintain-aspect-ratio
TRUE;display "ICON";valid_file "T";width 272.4375pt;height 210.4375pt;depth
0pt;original-width 582.75pt;original-height 449.0625pt;cropleft "0";croptop
"1";cropright "1";cropbottom "0";tempfilename
'figura1.wmf';tempfile-properties "NPR";}}

The SPDC quantum-injector is provided by a Type II $\Phi -phase\ tunable$
generator of linear polarization $({\bf \pi )}$-entangled photon couples.
The crystal realizing the OPA action is cut for Type II, noncollinear phase
matching and is equal to the one realizing SPDC. The detection system
consists of a birefringent plate $\Psi $, a ${\bf \pi }-rotator$ R($\varphi $%
), a polarizing-beam-splitter PBS and two cooled photomultipliers. In the
experiment a similar system is inserted on mode ${\bf k}_{2}$. We found that
the entanglement phase $\left| \Phi \right| $ of the output state of the
couple can be easily tuned over the range $0-\pi $ by rotating by an angle $%
\psi $ the crystal around the excitation axis ${\bf k}_{p}$, $\Phi (\psi )$
being a linear function \cite{branca}. In order to prevent any EPR type
state reduction that may affect the overall superposition process and then
destroy the S-cat at the outset, the photon \ emitted over the output mode $%
{\bf k}_{3}$ is filtered by a polarization analyzer with axis oriented at $%
45^{\circ }$to the horizontal (t.h.) before being detected by $D_{3}$ \cite
{epr}. An alternative solution for quantum injection, succesfully tested in
the experiment, is provided by a Type I NL crystal 1 feeding the OPA by a
single photon with ${\bf \pi }$ oriented at 45$^{\circ }$, the other photon
exciting $D_{3\text{ }}$without any ${\bf \pi -}selection$. In both cases, a
click at $D_{3}$ opens a gate selecting all registered outcomes, thus
providing the {\it conditional} character of the overall experiment . The
photon emitted over ${\bf k}_{1}$ provides the quantum-injection into the
OPA, physically consisting of the other NL crystal. The input state to our
amplifier system may be expressed in terms of superposition of Fock states
associated with the modes ${\bf k}_{j}(j=$ $1,2)$ and with the two ${\bf \pi 
}$ components respectively parallel and orthogonal (t.h.): 
\begin{equation}
\left| \Psi _{0}\right\rangle =2^{-%
{\frac12}%
}\left| 0\right\rangle _{2\perp }\otimes \left| 0\right\rangle _{2\mid \mid
}\otimes \left[ \left| 1\right\rangle _{1\perp }\otimes \left|
0\right\rangle _{1\mid \mid }+e^{i\Phi }\left| 0\right\rangle _{1\perp
}\otimes \left| 1\right\rangle _{1\mid \mid }\right]
\end{equation}
For a Type II NL\ crystal operating in noncollinear configuration the
overall amplification process taking place over ${\bf k}_{j}$ is contributed
by two equal and independent amplifiers $OPA_{A}$ and $OPA_{B}$ inducing
unitary transformations respectively on two couples of time dependent field
operators: $(\hat{a}_{1}(t)\equiv \hat{a}_{1\perp },\hat{a}_{2}(t)\equiv 
\hat{a}_{2\shortparallel })$ and $(\widehat{b}_{1}(t)\equiv \hat{a}_{1\mid
\mid },\widehat{b}_{2}(t)\equiv \hat{a}_{2\perp })$ for which, at the
initial time of the interaction, $t=0,$ is $\left[ \hat{a}_{i}(0),\hat{a}%
_{j}(0)^{\dagger }\right] =\left[ \widehat{b}_{i}(0),\widehat{b}%
_{j}(0)^{\dagger }\right] =\delta _{ij}$ and $\left[ \hat{a}_{i}(0),\widehat{%
b}_{j}(0)^{\dagger }\right] $ $=0$ for any $i$ and $j$ and $i,j=1,2$. A
quantum analysis of the dynamics of the system leads to a linear dependence
of the field operators on the corresponding input quantities, e.g. for $%
OPA_{A}$: 
\begin{equation}
\left[ 
\begin{array}{l}
\hat{a}_{1}(t) \\ 
\widehat{a}_{2}(t)^{\dagger }
\end{array}
\right] =\left[ 
\begin{array}{ll}
C & S \\ 
S & C
\end{array}
\right] \left[ 
\begin{array}{l}
\hat{a}_{1}(0) \\ 
\widehat{a}_{2}(0)^{\dagger }
\end{array}
\right]
\end{equation}
being: $C\equiv \cosh g,S$ $\equiv \sinh g,$ $g$ $\equiv \chi t$ the {\it %
amplification} {\it gain, }$\chi $ the coupling term proportional to the
product of the 2$^{nd}$-order NL susceptibility of the crystal and of the 
{\it pump} field, assumed classical and undepleted by the parametric
interaction. The interaction time, t may be determined in our case by the
length L of the NL crystal. The evolution operator for $OPA_{A}$ is then
expressed in the form of the unitary {\it squeeze operator}: $U_{A}(t)=$ $%
exp[g(\widehat{A^{\dagger }}-\widehat{A})]$ being: $\widehat{A^{\dagger }}%
\equiv \hat{a}_{1}(t)^{\dagger }\hat{a}_{2}(t)^{\dagger }$, $\widehat{A}%
\equiv \hat{a}_{1}(t)\hat{a}_{2}(t)$. A corresponding $U_{B}(t)$ for $%
OPA_{B} $ is given by the replacement $\hat{a}_{i}\rightarrow \widehat{b}%
_{i} $. By use of the overall propagator $U_{A}(t)U_{B}(t)$ and of the
disentangling theorem \cite{collett}, the output state is found: 
\begin{equation}
\left| \Psi \right\rangle \equiv G\left\{ \left| \Psi _{B}(0)\right\rangle
\otimes \left| \Psi _{A}(1)\right\rangle +e^{i\Phi }\left| \Psi
_{A}(0)\right\rangle \otimes \left| \Psi _{B}(1)\right\rangle \right\}
\end{equation}
where:$G\equiv (\sqrt{2}C^{2})^{-1}$;$\left| \Psi _{B}(0)\right\rangle
\equiv \sum\limits_{n=0}^{\infty }\sqrt{P_{n}}\left| n\right\rangle _{1\mid
\mid }\otimes \left| n\right\rangle _{2\perp }$, $\left| \Psi
_{A}(0)\right\rangle \equiv $ $\sum\limits_{n=0}^{\infty }\sqrt{P_{n}}$ $%
\left| n\right\rangle _{1\perp }\otimes \left| n\right\rangle _{2\mid \mid }$%
, $\Gamma \equiv S/C${\it \ }and $P_{n}\equiv $ $\overline{n}^{n}/(1+%
\overline{n})^{(1+n)}$= $\left( \Gamma ^{2n}/C^{2}\right) $ is a thermal{\it %
\ }distribution accounting for the squeezed-vacuum{\it \ }noise{\it \ }with
average photon number $\overline{n}=S^{2}$ \cite{walls}. Details on the
evaluation of (3) are given in Appendix A. below The two states espressed in
(3) as: $\left| \Psi _{A}(1)\right\rangle \equiv \sum\limits_{n=0}^{\infty
}\Gamma ^{n}\sqrt{n+1}\left| n+1\right\rangle _{1\perp }\otimes \left|
n\right\rangle _{2\mid \mid }$, $\left| \Psi _{B}(1)\right\rangle \equiv
\sum\limits_{n=0}^{\infty }\Gamma ^{n}\sqrt{n+1}\left| n+1\right\rangle
_{1\mid \mid }\otimes \left| n\right\rangle _{2\perp }$ represent the effect
of the one-photon quantum-injection{\it . }Since this sum is extended over
the complete set of n-states the appeal to the {\it macroscopic }quantum
coherence is justified{\it . }The output state function, written in the form 
$\left| \Psi \right\rangle =\left[ \left| \Psi _{\underline{A}}\right\rangle
+e^{+i\Phi }\left| \Psi _{\underline{B}}\right\rangle \right] $ with $\left|
\Psi _{\underline{A}}\right\rangle \equiv \left| \Psi _{B}(0)\right\rangle
\otimes \left| \Psi _{A}(1)\right\rangle $ and{\it \ }$\left| \Psi _{%
\underline{B}}\right\rangle \equiv \left| \Psi _{A}(0)\right\rangle \otimes
\left| \Psi _{B}(1)\right\rangle $ expresses the condition of quantum
superposition between two {\it pure,} multi-particle states originating,
through unitary OPA transformations, from the input single-particle state $%
\left| \Psi _{0}\right\rangle $, keeping in this process its original phase $%
\Phi $. In facts, any unitary transformation may generally transform but not
cancel the relevant quantum properties of the input state, such as
superposition and entanglement, even within a (noisy) process of particle
amplification as in our case. Furthermore, most important, since $\left|
\Psi \right\rangle $ is not factorizable in terms of linear polarization $%
{\bf \pi }-states,$ it keeps his original ${\bf \pi }-${\it entanglement}
character thus transferring into the multi-particle regime the striking
quantum nonseparability and Bell-type nonlocality properties of the
microscopic{\it \ (i.e. }2-particle) systems \cite{epr}\cite{su}. \ The $%
{\bf \pi -}${\it entanglement} properties of the output state can be
investigated experimetally either by a multi-particle Bell inequality
experiment, we shall consider in Section 5, or more simply and directly by
the {\it ad hoc} optical configuration already succesfully investigated in
our Laboratory for the case of a ${\bf \pi -}${\it entanglement} single
photon couple \cite{q-if}\cite{branca}.\ For convenience, a layout of a
possible experiment of this sort applied to our S-cat condition is reported
in the inset of Figure 1.

\section{Quantum-injected mode-degenerate OPA}

Consider now the diagram shown in Figure 2: the two NL\ crystals, e.g., BBO
cut for Type II phase-matching, are again excited by two beams derived from
the common UV\ source at wavelength $\lambda _{p}$. Crystal 1 is again the
SPDC\ source of couples of ${\bf \pi -}$entangled photons with $\lambda
=2\lambda _{p},$ emitted over the two output modes ${\bf k}_{i}$ $(i=1,2)$
determined by two fixed pinholes according to the phenomenology already
discussed in the previous Section.

Here again the photon emitted over ${\bf k}_{1}$provides the
quantum-injection into the OPA consisting of the NL crystal 2 which is now
cut for {\it collinear} operation over the two linear polarization modes $%
{\bf k}_{1\perp }$and ${\bf k}_{1\mid \mid }$, respectively parallel and
orthogonal t.h.. The photon associated with the output mode ${\bf k}_{2}$ is
filtered by a polarization analyzer with axis at $45^{\circ }$ respect to
the t.h.\ and then detected by $D_{2}$. Once again, the ${\bf \pi }-analyzer$
prevents any nonlocal, EPR type state reduction on the correlated mode ${\bf %
k}_{1}$and $D_{2}$ generates the gate pulse providing the {\it conditional}
character of the experiment. We may note that the input, quantum injection
single-photon state: 
\begin{equation}
\left| \Psi _{0}\right\rangle =2^{-%
{\frac12}%
}\left[ \left| 1\right\rangle _{1\perp }\otimes \left| 0\right\rangle
_{1\mid \mid }+(\exp i\Phi )\left| 0\right\rangle _{1\perp }\otimes \left|
1\right\rangle _{1\mid \mid }\right]
\end{equation}

is again entangled over the two polarization modes belonging to the common
momentum state ${\bf k}_{1}$. \ Note, by comparison with the input state
expressed by Equation 1, corresponding to the optical configuration 1, that
here two channels feeding vacuum into the NL crystal 2 are missing, since in
the present {\it collinear} configuration the $OPA_{A}$ and $OPA_{B}$
collapse into a single $OPA\ $and the dynamics is simplified. As we shall
see in Section 5 this results in a larger signal-to-noise ratio of the
output field. In order to investigate the properties of the output beam
after filtering against UV, a birefringent plate $D(\Psi )$ and a
Fresnel-rhomb $R(\varphi )$ induce respectively a field's phase delay $\Psi
=(\psi _{\perp }-\psi _{\mid \mid })$ and a $(45^{\circ }+\varphi $) ${\bf %
\pi -}rotation$ (t.h.). Then the two orthogonal ${\bf \pi }$ components are
detected by $D_{\varphi }$and $D_{\overline{\varphi }}$ after separation by
a polarizing beam splitter PBS. This realizes a ${\bf \pi -}$interferometer
as we shall see.

\FRAME{fhFU}{340.6875pt}{210.4375pt}{0pt}{\Qcb{Optical conf\`{i}guration of
the mode-degenerate entangled quantum injected OPA realizing the process of
multiphoton quantum superposition within a NL crystal cut for collinear type
II phase matching. For simplicity we keep the same denomination $k_{1}$ for
the injection mode involving both SPDC and OPA processes.}}{}{Figure }{%
\special{language "Scientific Word";type "GRAPHIC";maintain-aspect-ratio
TRUE;display "ICON";valid_file "T";width 340.6875pt;height 210.4375pt;depth
0pt;original-width 670.9375pt;original-height 412.75pt;cropleft "0";croptop
"1";cropright "1";cropbottom "0";tempfilename
'figura2.wmf';tempfile-properties "NPR";}}

We may analyze the present amplification process following the lines of the
theory given in Section 2. The single $OPA$ process induces a unitary
transformation on the couple of time dependent field operators: $\hat{a}%
(t)\equiv \hat{a}_{1\perp },$ and $\widehat{b}(t)\equiv \hat{a}_{1\mid \mid
} $ for which, at the initial time of the NL interaction, $t=0,$ is $\left[ 
\hat{a}(0),\hat{a}(0)^{\dagger }\right] =\left[ \widehat{b}(0),\widehat{b}%
(0)^{\dagger }\right] =1$ and $\left[ \hat{a}(0),\widehat{b}(0)^{\dagger }%
\right] $ $=0$. Note that the present notation for the denomination of the
field operators, is {\it not consistent} with the one adopted in Section 1.
It cannot lead to confusion but rather it is found convenient in view of the
comparative discussion, in the next Section, of the final expressions of
Wigner functions found for the two configurations. Here again quantum
analysis of the dynamics leads to a linear dependence of the field operators
on the corresponding input quantities: 
\begin{equation}
\left[ 
\begin{array}{l}
\hat{a}(t) \\ 
\widehat{b}(t)^{\dagger }
\end{array}
\right] =\left[ 
\begin{array}{ll}
C & S \\ 
S & C
\end{array}
\right] \left[ 
\begin{array}{l}
\hat{a}(0) \\ 
\widehat{b}(0)^{\dagger }
\end{array}
\right]
\end{equation}
The evolution operator is then expressed in the form of the {\it squeeze
operator}: $U_{A}(t)=$ $exp[g(\widehat{A^{\dagger }}-\widehat{A})]$ being: $%
\widehat{A^{\dagger }}\equiv \hat{a}(t)^{\dagger }\widehat{b}(t)^{\dagger }$,%
$\ \widehat{A}\equiv \hat{a}(t)\widehat{b}(t)$. The use of the {\it %
disentangling theorem}\cite{collett}, leads to the output state:

\begin{equation}
\left| \Psi \right\rangle _{OUT}=G\left[ \left| \overline{n}+1_{\perp
}\otimes \overline{n}_{\mid \mid }\right\rangle +\exp (i\Phi )\left| 
\overline{n}+1_{\mid \mid }\otimes \overline{n}_{\perp }\right\rangle \right]
\label{psiout}
\end{equation}
where $G\equiv (2C)^{-2}$. There the two mutually orthogonal, interfering 
{\it pure} states are now given in the form: 
\begin{equation}
\left| \overline{n}+1_{\perp }\otimes \overline{n}_{\mid \mid }\right\rangle
\equiv \sum\limits_{n=0}^{\infty }\Gamma ^{n}\sqrt{n+1}\left|
n+1\right\rangle _{1\perp }\otimes \left| n\right\rangle _{1\mid \mid }
\end{equation}
with $\Gamma \equiv S/C${\it . }It may be useful to express the output
function expressed by Equation 6 in an entangled form involving two
different output ${\bf k}-$vectors. In this connection a trivial example of
\ ``entanglement swapping'' may be easily realized by separating the two,
single momentum ${\bf k}_{1}$, linear polarization modes into two, single
polarization, momentum modes ${\bf k}_{3}$and ${\bf k}_{4}$, by a simple
insertion of a polarizing beam splitter (PBS)\ right at the output of the
OPA, as shown in Figure 2. In that case the state at the output of PBS is
given by:

\[
\left| \Psi \right\rangle _{OUT}=G\left[ \left| \overline{n}+1_{\perp
3}\otimes \overline{n}_{\mid \mid 4}\right\rangle +\exp (i\Phi )\left| 
\overline{n}_{\perp 3}\otimes \overline{n}+1_{\mid \mid 4}\right\rangle %
\right] 
\]

\begin{equation}
\left| \overline{n}+1_{\perp 3}\otimes \overline{n}_{\mid \mid
4}\right\rangle \equiv \sum\limits_{n=0}^{\infty }\Gamma ^{n}\sqrt{n+1}%
\left| n+1\right\rangle _{3\perp }\otimes \left| n\right\rangle _{4\mid \mid
}
\end{equation}
Note that the present entangled wavefunction is quite different from the one
given by Equation 3 and related to the optical configuration (1). The
entanglement character of the above function can be revealed by the same
experimental methods referred to at the end of Section 1.\FRAME{fhFU}{%
302.6875pt}{210.4375pt}{0pt}{\Qcb{Three-crystal variant of the mode
non-degenerate entangled OPA\ configuration. }}{}{Figure }{\special{language
"Scientific Word";type "GRAPHIC";maintain-aspect-ratio TRUE;display
"ICON";valid_file "T";width 302.6875pt;height 210.4375pt;depth
0pt;original-width 631.1875pt;original-height 437.4375pt;cropleft
"0";croptop "1";cropright "1";cropbottom "0";tempfilename
'figura3.wmf';tempfile-properties "NPR";}}

The optical configurations presented in Figures 1 and 2 are not the only
possible nor the more convenient ones. Consider for instance that the
mode-degenerate, collinear optical configuration shown in Figure 2 can be
easily replaced by a less elegant, {\it non-collinear} configuration in
which the OPA, consisting of a Type II NL crystal, is fed by a quantum
injection single-photon over two different input ${\bf k}_{j}$-vectors, say $%
j=1,\ 2$, each ${\bf k}_{j}$-vector corresponding to either one of the two
mutually orthogonal linear polarizations $\pi _{j}.$ \ The input
single-photon state is then: $\left| \Psi _{0}\right\rangle \ \ =\ 2^{-%
{\frac12}%
}\left[ \left| 1\right\rangle _{1\perp }\otimes \left| 0\right\rangle
_{2\mid \mid }+(\exp i\Phi )\left| 0\right\rangle _{1\perp }\otimes \left|
1\right\rangle _{2\mid \mid }\right] $. It is easy to recognize that, in
this case, we are led to the {\it same} results given by Equations 8 with
the output mode labels 3 and 4 replaced by 1 and 2. Alternatively, the OPA
may consist of a Type I crystal fed by quantum injection over the two ${\bf k%
}_{j}$-vectors with {\it equal} polarizations $\pi _{j}$ . In this case the
input state is simply expressed by $\left| \Psi _{0}\right\rangle \ \ =\ 2^{-%
{\frac12}%
}\left[ \left| 1\right\rangle _{1}\otimes \left| 0\right\rangle _{2}+(\exp
i\Phi )\left| 0\right\rangle _{1}\otimes \left| 1\right\rangle _{2}\right] \ 
$and again the output state is given by Equations 8.

In addition, consider the optical configuration presented in Figure 3. It is
a relevant three crystal variant of the above schemes and implies a {\it %
double quantum injection} into the OPA{\it \ , }e.g. by adoption of two
distinct SPDC processes feeding in {\it entangled-state} the two input modes
of the common entangled amplifier, ${\bf k}_{j}$ $(j=1,2)\ $within a {\it %
double-conditional}.experiment. In spite of the increased experimental
complications, mainly due to the low probability of the simultaneous OPA
quantum-injection processes, this new configuration may present definite
advantages. For instance, the {\it signal to noise }ratio\ and then the {\it %
visibility} $V$ of the Schroedinger-cat field are far larger that the one
for the mode non-degenerate OPA, as we shall see in the next Sections. In
addition and most important, the feeding if the common entangled OPA by two 
{\it simultaneous} quantum-injection processes with different phases $\Phi
_{j}$ adds new interesting quantum features to the output state of the
emitted field. \ We postpone to an {\it ad hoc} paper an exhaustive analysis
of such interesting complex optical configuration.

\section{ Wigner Function}

In order to inspect at a deeper lever the above results, consider the Wigner
function of the output field for the more complex configuration 1, shown in
Figure 1. Evaluate first the symmetrically-ordered characteristic function
of the set of complex variables $(\eta _{j},\eta \QATOP{\ast }{j},\xi
_{j},\xi \QATOP{\ast }{j})\equiv \left\{ \eta ,\xi \right\} $, $(j=1,2)$: $%
\chi _{_{S}}\left\{ \eta ,\xi \right\} \equiv \left\langle \Psi _{0}\right|
D[\eta _{1}(t)]D[\eta _{2}(t)]D[\xi _{1}(t)]D[\xi _{2}(t)]\left| \Psi
_{0}\right\rangle $ expressed in terms of the {\it displacement }operators: $%
D[\eta _{j}(t)]\equiv $ $\exp [\eta _{j}(t)\hat{a}_{j}(0)^{\dagger }-\eta
_{j}^{\ast }(t)\hat{a}_{j}(0)]$, ${\it \ }D[\xi _{j}(t)]$ $\ \equiv \exp
[\xi _{j}(t)\widehat{b}_{j}(0)^{\dagger }$ - $\xi _{j}^{\ast }(t)\widehat{b}%
_{j}(0)]$ \ where: $\eta _{1}(t)\equiv (\eta _{1}C-\eta \QATOP{\ast }{2}S)$; 
$\eta _{2}(t)\equiv (\eta _{2}C-\eta \QATOP{\ast }{1}S)$; $\xi _{1}(t)\equiv
(\xi _{1}C-\xi \QATOP{\ast }{2}S)$; $\xi _{2}(t)\equiv (\xi _{2}C-\xi \QATOP{%
\ast }{1}S)$. The Wigner function, expressed in terms of the corresponding
complex phase-space variables $(\alpha _{j}$, $\alpha _{j}^{\ast }$, $\beta
_{j}$, $\beta _{j}^{\ast })\equiv \left\{ \alpha ,\beta \right\} $ is the
eight-dimensional Fourier transform of $\chi _{_{S}}\left\{ \eta ,\xi
\right\} $, namely:

\begin{eqnarray}
W\left\{ \alpha ,\beta \right\} \text{{}} &=&\text{{}}\pi ^{-8}\iiiint
d^{2}\eta _{1}d^{2}\eta _{2}d^{2}\xi _{1}d^{2}\xi _{2}\chi _{_{S}}\left\{
\eta ,\xi \right\}  \nonumber \\
&&\ \ \ \ \ \ \ \ \ \ \exp \left\{ \sum_{j}[\eta \QATOP{\ast }{j}\alpha
_{j}-\eta _{j}\alpha \QATOP{\ast }{j}+\xi \QATOP{\ast }{j}\beta _{j}-\xi
_{j}\beta \QATOP{\ast }{j}]\right\}
\end{eqnarray}
where $d^{2}\eta _{j}\equiv d\eta _{j}d\eta \QATOP{\ast }{j}$, etc.$~$By a
lengthy application of operator algebra and integral calculus we could
evaluate analytically in closed form either $\chi _{_{S}}\left\{ \eta ,\xi
\right\} $\ and $W\left\{ \alpha ,\beta \right\} $.{\it \ }The corresponding
detailed calculations for the optical configuration (1) are reported in the
Appendix B and C. The final {\it exact} expression of the Wigner function
is: 
\begin{equation}
W\left\{ \alpha ,\beta \right\} =-\overline{W_{A}}\left\{ \alpha \right\} 
{\it \ }\overline{W_{B}}\left\{ \beta \right\} \left[ 1-\left| e^{i\Phi
}\Delta _{A}\left\{ \alpha \right\} +\Delta _{B}\left\{ \beta \right\}
\right| ^{2}\right]
\end{equation}
where $\Delta _{A}\left\{ \alpha \right\} \equiv 2^{-%
{\frac12}%
}(\gamma _{A+}-i\gamma _{A-})$, $\Delta _{B}\left\{ \beta \right\} \equiv
2^{-%
{\frac12}%
}(\gamma _{B+}-i\gamma _{B-})$ are expressed in terms of the squeezed
variables{\it : }$\gamma _{A+}\equiv (\alpha _{1}+\alpha _{2}^{\ast })e^{-g}$%
; $\gamma _{A-}\equiv i(\alpha _{1}-\alpha _{2}^{\ast })e^{+g}$; $\gamma
_{B+}\equiv (\beta _{1}+\beta _{2}^{\ast })e^{-g}$; $\gamma _{B-}\equiv
i(\beta _{1}-\beta _{2}^{\ast })e^{+g}$. The Wigner functions $\overline{%
W_{A}}\left\{ \alpha \right\} \equiv 4\pi ^{-2}\exp \left( -\left[ \left|
\gamma _{A+}\right| ^{2}+\left| \gamma _{A-}\right| ^{2}\right] \right) $; $%
\overline{W_{B}}\left\{ \beta \right\} \equiv 4\pi ^{-2}\exp \left( -\left[
\left| \gamma _{B+}\right| ^{2}+\left| \gamma _{B-}\right| ^{2}\right]
\right) $ definite positive over the 4 - dimensional spaces $\left\{ \alpha
\right\} $ and $\left\{ \beta \right\} $represent the effect of
squeezed-vacuum, i.e. emitted respectively by OPA$_{A}$\ and OPA$_{B}$ in
absence of any injection. Inspection of Equation 10 shows that precisely the
quantum superposition character of the injected state $\left| \Psi
_{0}\right\rangle $ determines the dynamical quantum superposition of the
devices $OPA_{A}$and $OPA_{B},$~the ones that otherwise act as {\it uncoupled%
} and{\it \ independent\ }objects. From another perspective, since the
quasi-probabilty functions $\overline{W_{A}}\left\{ \alpha \right\} $,{\it \ 
}$\overline{W_{B}}${\it \ }$\left\{ \beta \right\} $corresponding to the two
macrostates $\left| \Psi _{\underline{A}}\right\rangle \ $and $\left| \Psi _{%
\underline{B}}\right\rangle \ $in absence of quantum superposition are
defined in two totally separated and independent spaces, their respective
''distance'' in the overall phase-space of the system $\left\{ \alpha ,\beta
\right\} $ can be thought of as ''macroscopic'', as generally required by
any standard S-cat dynamics in a 2 - dimensional phase-space \cite{caldeira}%
. The link between the spaces $\left\{ \alpha \right\} $ and$\left\{ \beta
\right\} $ is provided by the quantum superposition term in Equation 10 $%
2Re[e^{i\Phi }\Delta _{A}\left\{ \alpha \right\} \Delta _{B}^{\ast }\left\{
\beta \right\} ]$. This term provides precisely the first-order quantum
interference of the macrostates $\left| \Psi _{\underline{A}}\right\rangle \ 
$and $\left| \Psi _{\underline{B}}\right\rangle $. In addition, and most
important, Equation 10 and Figure 3 show the non definite positivity{\it \ }%
of $W\left\{ \alpha ,\beta \right\} $ over its definition space. This
assures the overall quantum character of our multiparticle, quantum -
injected amplification scheme \cite{walls}\cite{gardiner}.

We may recognize that these last properties of the overall Wigner function
of our system do indeed coincide with the formal requirements of any
Schroedinger-cat apparatus, which may be outlined as follows \cite{zurek}:

(a) The ability of the system to create a first-order interference fringe
pattern is a necessary but{\it \ not sufficient} condition for any authentic
S-cat behaviour. The following are indeed the only two {\it necessary\ and} 
{\it sufficient} conditions.

(b) The Wigner function defined in the overall {\it phase-space} of the
system {\it must not }be definite-positive on his definition domain \cite
{walls}.

(c) The two interfering macrostates,\ identified \ by two \ corresponding \
gaussian-like peaks \ of \ the Wigner function must be clearly
distinguishable, i.e., the {\it distance} between the peaks must be larger
than their average width.

Note that conditions (b) and (c) imply necessarily the system's ability to
produce a first-order interference pattern while the inverse argument is not
necessarily true, as said. All these formal S-cat properties are shown by
the tridimensional plots of the reduced Wigner functions given in Figure 4.
These are drawn for $g=2.5$ and for different values of the injection phase $%
\Phi $, in correspondence with the configuration shown in Figure 1 and
investigated over the output mode ${\bf k}_{2}$ by the detection system
shown in the inset of the same Figure.

The Wigner function of the output field related to the quantum-injected {\it %
mode-degenerate OPA} considered in Figure 2 and Section 3 may be obtained by
a similar theoretical analysis. \ In facts this one is simpler because the 
{\it collinear} optical configuration adopted within the NL interaction
makes the two different amplifiers $A$ and $B$ of the previous case to
collapse into one. It follows that the dynamics of the system, instead of
being described in an eight dimensional phase-space, as in the previous
case, is described here in a more handy {\it four dimensional }space, i.e.
identified by two orthogonal ${\bf \pi \ }$states of the field emitted over
a {\it single} output mode, ${\bf k}_{1}$.

For the sake of completeness let's outline the theory in the {\it collinear}
case. Assume the {\it phase-space variables} $(\alpha ,\alpha ^{\ast },\beta
,\beta ^{\ast })\equiv \left\{ \alpha ,\beta \right\} $, the conjugated
variables $(\eta ,\eta ^{\ast },\xi ,\xi ^{\ast })\equiv \left\{ \eta ,\xi
\right\} $and evaluate the the {\it symmetrically }ordered {\it %
characteristic function}: $\chi _{_{S}}\left\{ \eta ,\xi \right\} $ $=$ $%
\left\langle \Psi _{0}\right| D[\eta (t)]D[\xi (t)]\left| \Psi
_{0}\right\rangle $ expressed in terms of the operators $D[\eta (t)]\equiv
\exp [\eta (t)\hat{a}(0)^{\dagger }-\eta ^{\ast }(t)\hat{a}(0)]$, $D[\xi
(t)]\equiv \exp [\xi (t)\widehat{b}(0)^{\dagger }-\xi ^{\ast }(t)\widehat{b}%
(0)]$ where $\eta (t)\equiv (\eta C-\eta ^{\ast }S)$; $\xi (t)\equiv (\xi
C-\xi ^{\ast }S)$ and $\hat{a}(0)$, $\widehat{b}(0)\ $are the field
operators associated to the two orthogonal input ${\bf \pi \ }$modes. The
Wigner function is the fourth dimensional Fourier transform of $\chi
_{S}\left\{ \eta ,\xi \right\} $.$~$

Once again, it is evaluated analytically in {\it closed form }and{\it \ }the
result is found to reproduce exactly the one expressed by Equation 10, after
a previous multiplication by $\pi ^{2}/4$. This unexpected result reached by
the analysis of two entirely different optical configurations suggests that
the actual form of $W\left\{ \alpha ,\beta \right\} $ given by Equation 10
is indeed determined by the peculiar character of the single-photon {\it %
quantum-injection} process which is common to both configurations. Of course
the parameters appearing in the expression of $W\left\{ \alpha ,\beta
\right\} $given by Equation 10 in correspondence with the {\it collinear}
case are now to be re-defined appropriately: $\gamma _{A+}\equiv (\alpha
+\beta ^{\ast })e^{-g}$, $\gamma _{A-}\equiv i(\alpha -\beta ^{\ast })e^{+g}$%
, $\gamma _{B+}\equiv (\beta +\alpha ^{\ast })e^{-g}$, $\gamma _{B-}\equiv
i(\beta -\alpha ^{\ast })e^{+g}$.

\FRAME{dtbpF}{392.0625pt}{188.3125pt}{0pt}{}{}{Figure }{\special{language
"Scientific Word";type "GRAPHIC";maintain-aspect-ratio TRUE;display
"ICON";valid_file "T";width 392.0625pt;height 188.3125pt;depth
0pt;original-width 620.75pt;original-height 296.75pt;cropleft "0";croptop
"1";cropright "1";cropbottom "0";tempfilename
'wigner1.wmf';tempfile-properties "NPR";}}

\FRAME{fhFU}{202.25pt}{174.6875pt}{0pt}{\Qcb{Tridimensional plots of the
Wigner function of the amplified field on mode ${\bf k}_{2}\ $at the output
of the quantum{\it \ }injected mode non-degenerate OPA as function of the 
{\it squeezed} variables: $X$= $(\protect\alpha +\protect\beta ^{\star
})e^{-g}$; $Y$= $i(\protect\beta -\protect\alpha ^{\star })e^{+g}$, for a
parametric gain g=2.5 and: $\Phi =0,\ \protect\pi /2,\ \protect\pi $.}}{}{%
Figure }{\special{language "Scientific Word";type
"GRAPHIC";maintain-aspect-ratio TRUE;display "ICON";valid_file "T";width
202.25pt;height 174.6875pt;depth 0pt;original-width 439.25pt;original-height
379.0625pt;cropleft "0";croptop "1";cropright "1";cropbottom
"0";tempfilename 'wigner2.wmf';tempfile-properties "NPR";}}

\section{Field-Correlation Functions}

Relevant information about the quantum mechanical features of the
quantum-injected OPA systems at hand are also revealed by the $1^{st}$ and $%
2^{nd}-order$ correlation functions of the output fields for both optical
configurations (1) and (2) corresponding repectively to the mode
non-degenerate and to the mode-degenerate cases \cite{walls}. Let's analyze
both cases in correspondence with the photo-detection measurements of the
output fields carried out by apparatuses equal to the one appearing in the
Inset of Figure1. Such measurement devices work as follows. Before detection
over the output momentum mode ${\bf k}_{j}$ $(j=1,2)$ the fields are
phase-shifted by $\Psi _{j}=(\psi _{j\perp }-\psi _{j\mid \mid })$ by a
birefringent plates and filtered by ${\bf \pi }$-analyzers with axes
oriented at the angles: $45^{0}+\varphi _{j}$ (t.h.). Each ${\bf \pi }$%
-analyzer may consist of the combination of a Fresnel-rhomb ${\bf \pi }$%
-rotator, $R(\varphi )$ and of a polarizing beam splitter, PBS: Figures 1
and 2. The field associated with the mode ${\bf k}_{j}$ is detected at the
space-time positions $x_{j}$ by two{\it \ linear }detectors $D_{j\varphi }$
and $D_{j\overline{\varphi }}$ with $\overline{\varphi }\equiv \varphi
+90^{0}$.

The $1^{st}-order$ correlation-functions $G\QATOP{(1)}{j}(x_{j},x_{j})\equiv
\left\langle \Psi _{0}\right| \widehat{N_{j}}(t)\left| \Psi
_{0}\right\rangle $ are ensemble averages of the number operators $\widehat{%
N_{j}}(t)\equiv \widehat{c}_{j}^{\dagger }(t)~\widehat{c}_{j}(t)$ written in
terms of the the detected output fields: $\widehat{c}_{j}(t)\equiv \lbrack
\xi _{j}^{-}~\hat{a}_{j}(t)+\xi _{j}^{+}~\widehat{b}_{j}(t)]$,$~[\widehat{c}%
_{i}(t),\widehat{c}\QATOP{\dagger }{j}(t)]=\delta _{ij}$, $\xi
_{j}^{+}\equiv 2^{-%
{\frac12}%
}(\cos \varphi _{j}+\sin \varphi _{j})\exp (i\psi _{j\beta })$, $\xi
_{j}^{-}\equiv 2^{-%
{\frac12}%
}(\cos \varphi _{j}-\sin \varphi _{j})\exp (i\psi _{j\alpha })$, where $\psi
_{j\alpha }$,$\psi _{j\beta }$ are phase-shifts induced by the birefringent
plate on the fields $\hat{a}_{j}(t)$, $\widehat{b}_{j}(t)$. These fields are
determined by the linear transformations given by Equations 2 and 5 for the
two configurations. In our cases $G\QATOP{(1)}{j}$ show the expected
superposition character of the output field with respect to the ${\bf \pi }$%
-rotation angles $\varphi _{j}$ and to the $\Delta \QATOP{\pm }{j}\Phi
\equiv (\Phi \pm \Psi _{j})$ being $\Phi $ the phase affecting the field's
output entangled state expressed by Equations 3 and 6. The explicit
evaluation of the first order functions for zero time delay leads to the
following results in correspondence with the optical configurations (1) and
(2).

The $2^{nd}-order$ functions$~G\QATOP{(2)}{ij}(x_{i},x_{j};x_{j,}x_{i})%
\equiv \left\langle \Psi _{0}\right| :\widehat{N_{i}}(t)\widehat{N_{j}}%
(t):\left| \Psi _{0}\right\rangle $ are also found for simultaneous
photo-detection processes on two equal or different {\bf k-}modes $i$,$j$ $%
(i,j=1,2)$.

\subsection{Mode non-degenerate OPA:}

The$1^{st}-order$ correlation functions are found to be expressed by: $G%
\QATOP{(1)}{1}=\overline{n}+%
{\frac12}%
(\overline{n}+1)[1+\cos (2\varphi _{1})\cos \Delta \QATOP{-}{1}\Phi ]$, $G%
\QATOP{(1)}{2}=$ $\overline{n}+%
{\frac12}%
$ $\overline{n}[1+\cos (2\varphi _{2})\cos \Delta \QATOP{+}{2}\Phi ]$.

These averages are related correspondingly to photodetection measurements
carried out over the output modes 1 or 2 by the detection apparatus shown in
the Inset of Figure 1. We may compare these results with the corresponding
averages over the input {\it vacuum state, }i.e., in.absence of any
quantum-injection process. These averages are found to be independent of $%
\varphi _{j}$ and $\Delta \Psi $, as expected and account for the
unavoidable, {\it squeezed} {\it vacuum} quantum noise affecting our active
parametric method: $G\QATOP{(1)}{1,vac}=G\QATOP{(1)}{2,vac}=\overline{n}$.
By this comparison we obtain the {\it signal-to-noise-ratio} related to the
Schroedinger-cat detection: $s/n$ $=$ $2,$ for $\Delta \QATOP{-}{j}\Phi
=\varphi _{j}=0$. The above result immediately suggests a $1^{st}$-$order~%
{\bf \pi -}$interferometric method for S-cat detection on a {\it single }$%
{\bf k}_{j}$ beam{\it ,} with {\it visibility: }${\it V}${\it \ = }$(G\QATOP{%
(1)}{\max }-G\QATOP{(1)}{\min })/(G\QATOP{(1)}{\max }+G\QATOP{(1)}{\min }%
)\geqq \frac{1}{3}$. The Wigner function plotted in Figure 4 refer to the
output field detected by this method on the mode ${\bf k}_{2}$. \ Note that
the average {\it difference} between the signals obtained at the output of
the detectors $D_{j\varphi }$ and $D_{j\overline{\varphi }}$ placed at the
output arms of PBS of the apparatus, Figure 1 inset, operating on the output
mode ${\bf k}_{j}$, expresses directly the {\it fringe pattern} related to
the $1^{st}-order$ interference between the two S-cat macrostates. For
instance for $j=2$, we have: $G\QATOP{(1)}{2}(\varphi _{2})-G\QATOP{(1)}{2}(%
\overline{\varphi }_{2})\ $= $\overline{n}\cos (2\varphi _{2})\cos \Delta 
\QATOP{+}{2}\Phi $. The $2^{nd}-order$ functions$~G\QATOP{(2)}{ij}%
(x_{i},x_{j};x_{j,}x_{i})\equiv \left\langle \Psi _{0}\right| :\widehat{N_{i}%
}(t)\widehat{N_{j}}(t):\left| \Psi _{0}\right\rangle $ are also found: $G%
\QATOP{(2)}{11}$= $2\overline{n}\left\{ \overline{n}\text{+}(\overline{n}%
+1)[1+\cos (2\varphi _{1})\cos \Delta \QATOP{-}{1}\Phi ]\right\} $; $G\QATOP{%
(2)}{22}=2\overline{n}^{2}\left\{ 1+\left[ 1\right. \right. $ $\left. \left.
+\cos (2\varphi _{2})\cos \Delta \QATOP{+}{2}\Phi \right] \right\} $; $G%
\QATOP{(2)}{12}$= $2\overline{n}^{2}$+$\overline{n}/2$+$\overline{n}[(%
\overline{n}+1)\cos (2\varphi _{1})\cos \Delta \QATOP{-}{1}\Phi )]$+ $%
\overline{n}(\overline{n}+1/2)\left[ 1\right. $ $\left. +\cos (2\varphi
_{2})\cos \Delta \QATOP{+}{2}\Phi \right] $ + $\overline{n}(\overline{n}%
+1)\left\{ [1\text{+}\cos \Delta \Psi ]\cos ^{2}\Delta \varphi ^{-}\right. $ 
$+$ $[1-\cos \Delta \Psi ]$ $\left. \sin ^{2}\Delta \varphi ^{+}\right\} $
where: $\Delta \varphi ^{\pm }\equiv (\varphi _{1}\pm \varphi _{2})$, $%
\Delta \Psi \equiv (\Psi _{1}+\Psi _{2})$.We may easily prove, e.g. for all $%
\Delta \QATOP{\pm }{j}\Phi $= $\varphi _{j}$=$0$,$~$that our system realizes
the {\it maximum} quantum mechanical violation of the Cauchy-Schwarz
inequality which generally holds in semi-classical field theory: $[g\QATOP{%
(2)}{12}(0)]^{2}\leq g\QATOP{(2)}{11}(0)~g\QATOP{(2)}{22}(0)$ being: $g%
\QATOP{(2)}{i~j}(0)\equiv G\QATOP{(2)}{i~j}(0)[G\QATOP{(1)}{i}(0)$ $G\QATOP{%
(1)}{j}(0)]^{-1}$\cite{walls}. Furthermore, the given expression of $G\QATOP{%
(2)}{12}$ shows the effects of the {\it multiparticle} quantum
nonseparability and Bell-type nonlocality, contributed by the terms
proportional to $\cos (\Delta \varphi ^{\pm })$ and $\cos \Delta \Psi $.
This is a most relevant manifestation of the nonlocality properties of our
quantum injected, {\it entangled} parametric system \cite{epr}\cite{su}.

\subsection{Mode-degenerate OPA}

The analysis above can be repeated for the simpler dynamics of the {\it %
collinear} case, configuration (2). The results are:

$1^{st}-order$ correlation function: $G^{(1)}(\varphi )=\overline{n}+(%
\overline{n}+%
{\frac12}%
)[1+\cos (2\varphi )\cos \Delta \Phi ]$, with $\Delta \Phi \equiv (\Psi
-\Phi )$. This leads to an {\it interference} {\it fringe visibility: }$%
V=(G_{max}^{(1)}-G_{min}^{(1)})/(G_{max}^{(1)}+G_{min}^{(1)})\geq 
{\frac12}%
,$ for $\Delta \Phi \equiv 0$. Note that in the collinear case $V$ is larger
by a factor $1.5$ respect to the noncollinerar case, a result due to the
absence in the dynamics of the input vacuum fields contributions over the
mode ${\bf k}_{1}$, Figure1. The absence of an input vacuum field on the
idler mode is a most favorable condition shared also by the three-crystal
OPA\ configuration shown in Figure 3. Again the {\it fringe pattern }related
to the 1$^{st}$ order interference between the two S-cat macrostates is
determined by the difference: $G^{(1)}(\varphi )-G^{(1)}(\overline{\varphi }%
)=\overline{n}\cos (2\varphi )\cos \Delta _{2}^{+}\Phi .$

$\ $

The $2^{nd}$-$order$ correlation functions $G\QATOP{(2)}{\varphi ,\varphi
^{\prime }}\equiv \left\langle \Psi _{0}\right| :[\widehat{N}(t)]_{\varphi }[%
\widehat{N}(t)]_{\varphi ^{\prime }}:\left| \Psi _{0}\right\rangle $ may
also be measured by use of the detectors $D_{\varphi }$ and $D_{\overline{%
\varphi }}$, Figure2$.$.Their expressions are given here for completeness: $G%
\QATOP{(2)}{\varphi ,\varphi }=$ $6\overline{n}^{2}\ $+$\ 2\overline{n}\ $+ $%
3\overline{n}(\overline{n}+1)\cos ^{2}(2\varphi )$ + $2\overline{n}(3%
\overline{n}+2)\cos (2\varphi )\cos \Delta \Phi $; $G\QATOP{(2)}{\varphi ,%
\overline{\varphi }}$= $2\overline{n}(3\overline{n}+2)$ - $3\overline{n}(%
\overline{n}+1)\cos ^{2}(2\varphi )$.

\section{Decoherence. Conclusions.}

The virtual absence of any effective decoherence within the travelling-wave
(TW)\ parametric process we are considering may be understood by the
following argument. Our Schroedinger-cat system does not consist of a {\it %
free }excitation{\it ,} a condition common to~{\it all~} S-cats considered
thus far in the literature. It consists in fact of a {\it driven} excitation
which it is strongly coupled with a {\it continuously} {\it re-phasing}
environment provided by the coherent{\it \ nonlinear }polarization of the
parametric process {\it driving\ }the multiphoton field in quantum
superposition. Similar situations of coherence persistence {\it in spite} of
dissipation are often encountered in physics of the {\it nonlinear }%
dynamical systems, e.g., in solid state nonlinear spectroscopy. A nice
example is provided there by the nonlinear generation of surface-polaritons
or plasmons in strongly light absorbing semiconductors or metals \cite
{demartini}. In spite of any arbitrarily large damping of the medium, high
intensity and {\it strictly} {\it nondecaying,} {\it driven} polariton waves%
{\it \ }may be{\it \ }nonlinearly{\it \ }generated over arbitrarily{\it \ }%
large distances and times while the related {\it free }waves{\it , }that
originate at the boundaries from the{\it \ }driven{\it \ }ones{\it , }die
out quickly according to the linear optical properties of the medium. Note
that in the {\it linear} regime, i.e. where the only driving force is the
linear polarization, the two kinds of waves coincide and are damped at the
same rate. We believe that the above interpretation is generally valid for
any kind of dissipative process, e.g. damping, de-phasing and de-coherence.
Since this one is a most disruptive process for quantum coherence in complex
systems, e.g. in the domain of {\it quantum computation}, our results would
then suggest the {\it nonlinear} interaction among the information carrying
particles as a most efficient solution toward a large scale implementation
of the new methods \cite{divince}. Of course any single photon loss event,
mainly contributed in the present TW case by stray reflections, implies an
elementary decoherence process. In our laboratory experiment two equal 1mm
thick, BBO crystals are excited by 0.8 picosecond pulses at $\lambda
_{p}=400nm$ second-harmonic-generated by a mode-locked Ti : Sa laser at a 76
MHz repetition rate with an average power $\approx 0.3W$. The detection
system, consisting of two linear photodetectors connected to an electronic
correlator, is equal to the one shown by Figure1 inset, but for the absence
of the birefringent plate. The initial phase is: $\Phi =0.$ All surfaces are
treated by special AR coatings resonant at the working $\lambda =800nm$ with
an overall transmittivity: $T\approx 99.60\%$. This figure implies the loss
of a single photon every $\gtrsim 20$ pulses with the generation of $%
\overline{n}\approx 10$ per pulse. This would make our S-cat experiment
quite feasible. Note in this connection that the{\it \ travelling-wave} case
is quite superior to the optical parametric {\it oscillator} (OPO)
configuration where the presence of unavoidable cavity losses enhance the
negative effect of all phase-disrupting processes \cite{Fdemartini}.

In summary, we have given the quantum analysis of a novel nonlinear
entangled TW parametric system that shows macroscopic, decoherence free,
quantum superposition features that can be easily detectable. This result is
reached by a smart interplay of the fundamental paradigms of modern quantum
optics, i.e., {\it quadrature squeezing},{\it \ multiparticle state
entanglement} and {\it quantum nonseparability }in parametric correlations.
From a foundational perspective, our method could find application within
the realization of fundamental nonlocality and noncontestuality tests of
quantum mechanics requiring a number of entangled particles larger than two 
\cite{mermin}. In addition, within the fields of {\it quantum information }%
and {\it computation} it may represent a new way to amplify quantum
coherence and entanglement over large systems providing at the same time an
elegant way to beat decoherence. Of course we are dealing here with a {\it %
noisy} system. However the ''quantum-noise-reduction effect'' contributed
efficiently by the parametric{\it \ quadrature squeezing} may find here a
successful application \cite{walls}.{\it \ }For many reasons we are inclined
to believe that, if successful, the present project may open new and long
reaching paths in some fundamental chapters of modern physics. We thank
S.Branca, M.D'Ariano, G.Di Giuseppe, D.P.Di\ Vincenzo, G.Ghirardi for
enlightening discussions, the CEE-TMR Program (Contract N.ERBMRXCT96-066)
and INFM (Contract PRA97-cat) for funding.\newpage

\section*{Appendix A: Output wavefunction}

The application of the {\it disentangling theorem }in the context of the
present work to the general input Fock state $(\left| n\right\rangle
_{1\perp }\otimes \left| m\right\rangle _{2\mid \mid })_{in}\ $implies the
use of the following transformations.

$exp[g(\widehat{A^{\dagger }}-\widehat{A})](\left| n\right\rangle _{1\perp
}\otimes \left| m\right\rangle _{2\mid \mid })_{in}=\exp \Gamma \widehat{%
A^{\dagger }}\ exp(\varsigma \lbrack \widehat{A^{\dagger }},\widehat{A}%
])\exp (-\Gamma \widehat{A})(\left| n\right\rangle _{1\perp }\otimes \left|
m\right\rangle _{2\mid \mid })_{in}$, where: $\varsigma \equiv \ln C$. \
Since $[\widehat{A^{\dagger }},\widehat{A}]=-\{\hat{a}_{1}(t)^{\dagger }\hat{%
a}_{1}(t)+\hat{a}_{2}(t)^{\dagger }\hat{a}_{2}(t)+1\}$ the following results
are found for three relevant input states.

$(a)$\ $exp[g(\widehat{A^{\dagger }}-\widehat{A})](\left| 0\right\rangle
_{1\perp }\otimes \left| 0\right\rangle _{2\mid \mid })=exp\{-\varsigma
+\Gamma \widehat{A^{\dagger }}\}(\left| 0\right\rangle _{1\perp }\otimes
\left| 0\right\rangle _{2\mid \mid })\ =\sum \QATOP{\infty }{n=0}\sqrt{P_{n}}%
\left| n\right\rangle _{1\perp }\otimes \left| n\right\rangle _{2\mid \mid
}= $ $\left| \Psi _{B}(0)\right\rangle $, where $P_{n}\equiv C^{-2}\Gamma
^{2n}=\overline{n}^{n}/(\overline{n}+1)^{n+1}$ expresses the {\it thermal
distribution} of the squeezed-vacuum on the two output modes with average
photon number: $\overline{n}=S^{2}$.

$(b)\ exp[g(\widehat{A^{\dagger }}-\widehat{A})](\left| 1\right\rangle
_{1\perp }\otimes \left| 0\right\rangle _{2\mid \mid })=\exp \Gamma \widehat{%
A^{\dagger }}\ exp(-\varsigma \lbrack \widehat{A^{\dagger }},\widehat{A}%
])(\left| 1\right\rangle _{1\perp }\otimes \left| 0\right\rangle _{2\mid
\mid })\ =$ $C^{-2}\sum \QATOP{\infty }{n=0}\frac{\Gamma ^{n}}{n!}(\hat{a}%
_{1}^{\dagger })^{n+1}(\hat{a}_{2}^{\dagger })^{n}(\left| 0\right\rangle
_{1\perp }\otimes \left| 0\right\rangle _{2\mid \mid })\ =$\ $C^{-2}\sum 
\QATOP{\infty }{n=0}\Gamma ^{n}\sqrt{n+1}(\left| n+1\right\rangle _{1\perp
}\otimes \left| n\right\rangle _{2\mid \mid })\ =\left| \Psi
_{A}(1)\right\rangle .$

$(c)\ exp[g(\widehat{A^{\dagger }}-\widehat{A})](\left| 1\right\rangle
_{1\perp }\otimes \left| 1\right\rangle _{2\mid \mid })=\exp (-3\varsigma
)\sum \QATOP{\infty }{n=0}\frac{\Gamma ^{n}}{n!}(\hat{a}_{1}^{\dagger }\hat{a%
}_{2}^{\dagger })^{n}(\left| 1\right\rangle _{1\perp }\otimes \left|
1\right\rangle _{2\mid \mid })$- $\exp (-\varsigma )\Gamma \sum \QATOP{%
\infty }{n=0}\Gamma ^{n}(\left| n\right\rangle _{1\perp }\otimes \left|
n\right\rangle _{2\mid \mid })\ =C^{-3}\sum \QATOP{\infty }{n=0}\Gamma
^{n}(n+1)(\left| n\right\rangle _{1\perp }\otimes \left| n\right\rangle
_{2\mid \mid })$-$\Gamma \left| \Psi _{B}(0)\right\rangle $. The other
states appearing in Equation \ 3 are obtained by the same transformations
upon the substitutions: A$\leftrightarrows B,$ $\Vert \leftrightarrows \bot
\ $etc.

\section*{Appendix B: Characteristic function\protect\bigskip}

The symmetrically ordered characteristic\ function or the optical
configuration (1) is evaluated by\ the average:

$\chi _{_{S}}\left\{ \eta ,\xi \right\} \equiv $ $\left\langle \Psi
_{0}\right| D[\eta _{1}(t)]D[\eta _{2}(t)]D[\xi _{1}(t)]D[\xi _{2}(t)]\left|
\Psi _{0}\right\rangle $ expressed in terms of the displacement operators: $%
\ D[\eta _{j}(t)]$ \ \ \ $\equiv \exp [\eta _{j}(t)\hat{a}_{j}(0)^{\dagger
}-\eta _{j}^{\ast }(t)\hat{a}_{j}(0)]$; $D[\xi _{j}(t)]\equiv \exp [\xi
_{j}(t)\widehat{b}_{j}(0)^{\dagger }-\xi _{j}^{\ast }(t)\widehat{b}_{j}(0)]$
and of the time dependent parameters: $\eta _{1}(t)\equiv (\eta _{1}C-\eta 
\QATOP{\ast }{2}S)$; $\eta _{2}(t)\equiv (\eta _{2}C-\eta \QATOP{\ast }{1}S)$%
; $\xi _{1}(t)\equiv (\xi _{1}C-\xi \QATOP{\ast }{2}S)$; $\xi _{2}(t)\equiv
(\xi _{2}C-\xi \QATOP{\ast }{1}S)$ where $(\eta _{j},\eta \QATOP{\ast }{j}%
,\xi _{j},\xi \QATOP{\ast }{j})\equiv \left\{ \eta ,\xi \right\} $, $(j$=$%
1,2)$ represents the set of the eight phase-space conjugate complex
variables relative to our dynamical problem . Note that the expression of $%
\chi _{_{S}}\left\{ \eta ,\xi \right\} $ may be given in an equivalent,
alternative form by use of the following results: $D[\eta _{1}(t)]D[\eta
_{2}(t)]=D[\eta _{1}]D[\eta _{2}]$ and $D[\xi _{1}(t)]D[\xi _{2}(t)]=D[\xi
_{1}]D[\xi _{2}]$ where: $D[\eta _{j}]\equiv \exp [\eta _{j}\hat{a}%
_{j}(t)^{\dagger }-\eta _{j}^{\ast }\hat{a}_{j}(t)]$;\ $D[\xi _{j}]\equiv
\exp [\xi _{j}\widehat{b}_{j}(t)^{\dagger }-\xi _{j}^{\ast }\widehat{b}%
_{j}(t)]$ These last results are obtained by use of Eqs.2 and of the well
known theorem: $\exp \widehat{A}$ $exp\widehat{B}$ $=$ $\exp (\widehat{A}+%
\widehat{B})\exp 
{\frac12}%
\left[ \widehat{A},\widehat{B}\right] $. We may then evaluate the explicit
expression of $\chi _{_{S}}\left\{ \eta ,\xi \right\} $ by use of the
explicit form of the input state Eq.1 and of the well known relations
involving {\it displacement operators :} $D^{\dagger }(\alpha
)=D^{-1}(\alpha )=D(-\alpha )$, $D^{\dagger }(\alpha )\widehat{a}D(\alpha )=%
\widehat{a}+\alpha $, $D^{\dagger }(\alpha )\widehat{a}^{\dagger }D(\alpha )=%
\widehat{a}^{\dagger }+\alpha ^{\ast }$, $<0\mid D(\alpha )\mid 0>=<0\mid
\alpha >=\exp (-%
{\frac12}%
\mid \alpha \mid ^{2})$ \cite{walls}. By ensemble averaging over the two
superposition terms appearing in Eq.1, the symmetrically ordered
characteristic function is finally found: 
\begin{equation}
\chi _{_{S}}\left\{ \eta ,\xi \right\} =\left\{ 1-%
{\frac12}%
\mid e^{i\Phi }\eta _{1}(t)+\xi _{1}(t)\mid ^{2}\right\} \exp [-%
{\frac12}%
\sum_{j}(\mid \eta _{j}(t)\mid ^{2}+\mid \xi _{j}(t)\mid ^{2})]
\end{equation}

\bigskip

\section*{Appendix C: Wigner Function}

In view of the explicit form of $\chi _{_{S}}\left\{ \eta ,\xi \right\} \ $%
the $8^{th}$-dimensional integral, Equation 9 is evaluated by introducing
first the change of variables: $\eta _{j}\rightarrow \eta _{j}(t)$, $\xi
_{j}\rightarrow \xi _{j}(t)\ $and then by writing:$\ \eta _{j}(t)\equiv
x_{j}e^{i\varphi _{j}}$, $\xi _{j}(t)\equiv y_{j}e^{i\overline{\varphi }%
_{j}} $, $d^{2}\eta _{j}\equiv d\eta _{j}(t)d\eta \QATOP{\ast }{j}%
(t)=x_{j}dx_{j}d\varphi _{j}$;$\ d^{2}\xi _{j}(t)\equiv d\xi _{j}(t)d\xi 
\QATOP{\ast }{j}(t)=y_{j}dy_{j}d\overline{\varphi }_{j}\ $where $\left| \eta
_{j}(t)\right| \equiv x_{j}(t)\equiv x_{j}\ $and$\ \left| \xi _{j}(t)\right|
\equiv y_{j}(t)\equiv y_{j}$, $j=1,2$. \ For integration purposes this
transformation is completed by multiplication of the integrand by the
determinant of the $8\times 8\ $Jacobian matrix,$\left[ D_{ij}\right] \ $%
with elements: $D_{ij}=\partial \left\{ \eta ,\xi \right\} _{i}/\partial
\left\{ \eta (t),\xi (t)\right\} _{j}\ $with obvious notation for the
partial derivatives. It is convenient to re-define here, by a multiplication
by $\sqrt{2}$, the linear transformations between the two sets of variables,
i.e.: $\eta _{1}=\sqrt{2}[\eta _{1}(t)C+\eta \QATOP{\ast }{2}(t)S]$; $\eta
_{2}=\sqrt{2}[\eta _{2}(t)C+\eta \QATOP{\ast }{1}(t)S]$; $\xi _{1}=\sqrt{2}[%
\xi _{1}(t)C+\xi \QATOP{\ast }{2}(t)S]$; $\xi _{2}=\sqrt{2}[\xi _{2}(t)C+\xi 
\QATOP{\ast }{1}(t)S]$. The value of the Jacobian determinant is then found
=16. By the above substitutions, the argument of the exponential in the
integrand of the integral, Eq.9 may be cast in the form: $\sum_{j}[\eta 
\QATOP{\ast }{j}\alpha _{j}-\eta _{j}\alpha \QATOP{\ast }{j}+\xi \QATOP{\ast 
}{j}\beta _{j}-\xi _{j}\beta \QATOP{\ast }{j}]\ =\ \sum_{j}[p_{j}\cos
\varphi _{j}+q_{j}\sin \varphi _{j}+\overline{p}_{j}\cos \overline{\varphi }%
_{j}+\overline{q}_{j}\sin \overline{\varphi }_{j}]$, where: $p_{1}=\sqrt{2}%
x_{1}[S(\alpha _{2}-\alpha \QATOP{\ast }{2})+C(\alpha _{1}-\alpha \QATOP{%
\ast }{1})]$; $q_{1}=i\sqrt{2}x_{1}[S(\alpha _{2}+\alpha \QATOP{\ast }{2}%
)-C(\alpha _{1}+\alpha \QATOP{\ast }{1})]$ . The couple $(p_{2},\ q_{2})\ $%
is obtained by entering the substitutions: $\alpha _{1}\leftrightarrows
\alpha _{2}\ $and $x_{1}\rightarrow x_{2}$ in the expressions for $(p_{1},\
q_{1})$. Likewise, the couples $(\overline{p}_{1},\ \overline{q}_{1})$ and $(%
\overline{p}_{2},\ \overline{q}_{2})\ $are found by the substitutions: $%
\alpha _{j}\rightarrow \beta _{j},x_{j}\rightarrow y_{j}\ $in the
expressions for $(p_{1,}\ q_{1})$ and $(p_{2,}\ q_{2})$. Consider now the
following expressions: $w_{j}\equiv \sqrt{p\QATOP{2}{j}+q\QATOP{2}{j}}$=$%
ix_{j}\left| \delta _{j}\right| $, $\overline{w}_{j}\equiv \sqrt{\overline{p}%
\QATOP{2}{j}+\overline{q}\QATOP{2}{j}}=i\ y_{j}\left| \overline{\delta }%
_{j}\right| $, where: $\delta _{1}\equiv 2\sqrt{2}[C\alpha _{1}-S\alpha 
\QATOP{\ast }{2}]$, $\delta _{2}\equiv 2\sqrt{2}[C\alpha _{2}-S\alpha \QATOP{%
\ast }{1}]$ and $\overline{\delta }_{j}$ is obtained again by the
substitutions: $\alpha _{j}\rightarrow \beta _{j}$. Turn now to Equation 9
and integrate first respect to the phases $\varphi _{j}$, $\overline{\varphi 
}_{j}$. This step is accomplished by the use of the following results:

$(a)$ $\int \QATOP{2\pi }{0}\exp (p_{j}\cos \varphi _{j}+q_{j}\sin \varphi
_{j})\ d\varphi _{j}=2\pi {\bf J}_{0}(i$\bigskip $w_{j})$\ \ [GR 3.937,
pag.488; \cite{GR}].

$(b)\int \QATOP{2\pi }{0}\exp (p_{j}\cos \varphi _{j}+q_{j}\sin \varphi
_{j})\ \sin \varphi _{j}d\varphi _{j}=-i2\pi \frac{q_{j}}{w_{j}}{\bf J}%
_{1}(i $\bigskip $w_{j})$.

$(c)\int \QATOP{2\pi }{0}\exp (p_{j}\cos \varphi _{j}+q_{j}\sin \varphi
_{j})\ \cos \varphi _{j}d\varphi _{j}=$ $-i2\pi \frac{p_{j}}{w_{j}}{\bf J}%
_{1}(i$\bigskip $w_{j})$, \ being ${\bf J}_{\nu }(z)$ a Bessel function.

We may now integrate respect to $x_{j}$ and $y_{j}$. For this purpose let's
first evaluate the following integral involving some relevant expressions of
the regular function $F(x_{j},y_{j})$:

\begin{eqnarray*}
I_{W}[F(x_{j},y_{j})]\ &=&\ \pi ^{-8}\iiiint d^{2}\eta _{1}d^{2}\eta
_{2}d^{2}\xi _{1}d^{2}\xi _{2}F(x_{j},y_{j}) \\
&&\ \ \ \ \ \ \ \ \ \ \ \ \ \exp \left\{ \sum_{j}[(\eta \QATOP{\ast }{j}%
\alpha _{j}-\eta _{j}\alpha \QATOP{\ast }{j})+(\xi \QATOP{\ast }{j}\beta
_{j}-\xi _{j}\beta \QATOP{\ast }{j})]\right\}
\end{eqnarray*}
$.$

$(d)\ $Let $F(x_{j},y_{j})=\exp [-\sum_{j}(x\QATOP{2}{j}+y\QATOP{2}{j})]$
and consider the integral [GR 4, pag. 717]: $\int \QATOP{\infty }{0}x^{\nu
+1}e^{-\alpha x^{2}}{\bf J}_{\nu }(\beta x)dx=\beta ^{\nu }(2\alpha )^{-(\nu
+1)}e^{-(\beta ^{2}/4\alpha )}$. Assuming $\nu =0$, $\alpha =1$, $\beta
=-\left| \delta \right| $ this result leads immediately to: $%
I_{W}[F(x_{j},y_{j})]=16\pi ^{-4}\exp [-%
{\frac14}%
\sum_{j}(\left| \delta _{j}\right| ^{2}+\left| \overline{\delta }_{j}\right|
^{2})]$.

$(e)\ $Let $F(x_{j},y_{j})=(x\QATOP{2}{j}+y\QATOP{2}{j})\exp [-\sum_{j}(x%
\QATOP{2}{j}+y\QATOP{2}{j})]$ and consider the integral [GR 6.631, pag.716]: 
$\int \QATOP{\infty }{0}x^{\mu }e^{-\alpha x^{2}}{\bf J}_{\nu }(\beta
x)dx=\beta ^{\nu }{\bf \Gamma }(\nu /2+\mu /2+%
{\frac12}%
)$ $[2^{\nu +1}\alpha ^{%
{\frac12}%
(\nu +\mu +1)}{\bf \Gamma }(\nu +1)]^{-1}{\bf \Phi }[%
{\frac12}%
(\nu +\mu +1),\ \nu +1;\ -\beta ^{2}/(4\alpha )]\ $being ${\bf \Phi }(\alpha
^{\prime },\gamma ^{\prime };z^{\prime })$ a {\it degenerate hypergeometric}
function. Apply the Kummer theorem: ${\bf \Phi }(\alpha ^{\prime },\gamma
^{\prime };z^{\prime })$ $\ =\ e^{z^{\prime }}{\bf \Phi }(\gamma ^{\prime
}-\alpha ^{\prime },\gamma ^{\prime };-z^{\prime })$ to the standard
infinite series expansion: ${\bf \Phi }(\alpha ^{\prime },\gamma ^{\prime
};z^{\prime })=$ $1+\frac{\alpha ^{\prime }z^{\prime }}{\gamma ^{\prime }1!}+%
\frac{\alpha ^{\prime }(\alpha ^{\prime }+1)z^{\prime 2}}{\gamma ^{\prime
}(\gamma ^{\prime }+1)2!}+..$[GR 9.210 and 9.212]. Note that with the values
of the parameters: $\nu =0$, $\mu =3$, $\alpha =1$, $\beta =-\left| \delta
\right| $ only the first two terms of the expansion are nonzero, i.e. ${\bf %
\Phi }(2,1;-%
{\frac14}%
\left| \delta \right| ^{2})$= $[1-%
{\frac14}%
\left| \delta \right| ^{2}]\exp (-%
{\frac14}%
\left| \delta \right| ^{2})$. This leads to the {\it exact} result: $%
I_{W}[F(x_{j},y_{j})]\ =16\pi ^{-4}[2-%
{\frac14}%
(\left| \delta _{j}\right| ^{2}+\left| \overline{\delta }_{j}\right|
^{2})\exp [-%
{\frac14}%
\sum_{j}(\left| \delta _{j}\right| ^{2}+\left| \overline{\delta }_{j}\right|
^{2})]$

$(f)\ $Let $F(x_{j},y_{j})=x_{j}\ y_{j}\exp i(\varphi _{j}-\overline{\varphi 
}_{j})\exp [-\sum_{j}(x\QATOP{2}{j}+y\QATOP{2}{j})]$ and consider the
integral GR 6.631 just given at paragraph $(e)$. With the new set of
parameters: $\nu =1$, $\mu =2$, $\alpha =1$, $\beta =-\left| \delta \right| $%
, and by use of the quoted Kummer theorem within the series expansion for $%
{\bf \Phi }(2,2;-%
{\frac14}%
\left| \delta \right| ^{2})$, the {\it exact} value of the integral is
found: $-%
{\frac14}%
\left| \delta \right| \exp (-%
{\frac14}%
\left| \delta \right| )$. This leads to the result: $(w_{j})^{-1}\int \QATOP{%
\infty }{0}x\QATOP{2}{j}J_{1}(-x_{j}\left| \delta _{j}\right|
)dx_{j}=i(4x_{j})^{-1}\exp (-%
{\frac14}%
\left| \delta _{j}\right| )$. A further calculation of an identical integral
involving the variables $y_{j}$, $\overline{w}_{j}$, $\overline{\delta }%
_{j}\ $leads to the result: $I_{W}[F(x_{j},y_{j})]\ =\ -(16\pi ^{-4})\delta
_{j}\overline{\delta }\QATOP{\ast }{j}\exp [-%
{\frac14}%
\sum_{j}(\left| \delta _{j}\right| ^{2}+\left| \overline{\delta }_{j}\right|
^{2})]$. \ At last we insert the explicit expression of $\chi _{_{S}}\left\{
\eta ,\xi \right\} $ found in Appendix B within the integral expressed by
Equation 9 and use the above results $d$, $e$, $f$. This leads to the final
expression of $W\left\{ \alpha ,\beta \right\} $ given by Equation 10.


\begin{thebibliography}{99}
\bibitem{divince}  The realization of nonlinear universal logic gates is now
being considered as a sensible solution toward experimental quantum
computation:\ D. Di Vincenzo (private comm.).
\end{thebibliography}
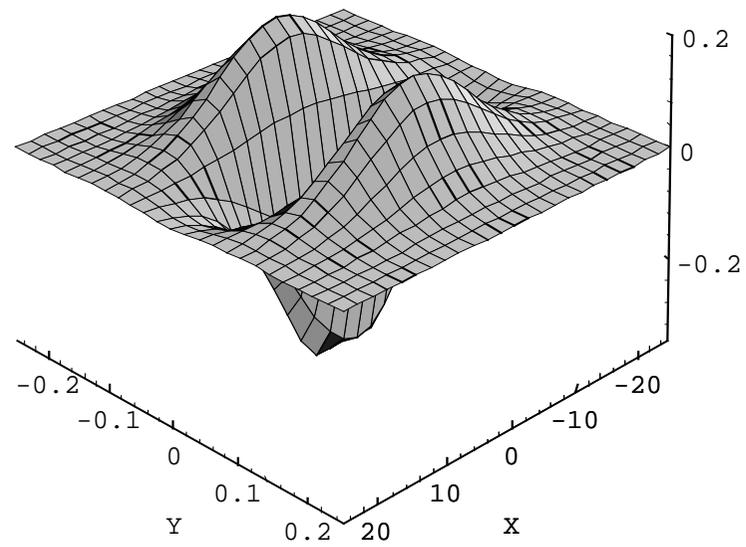
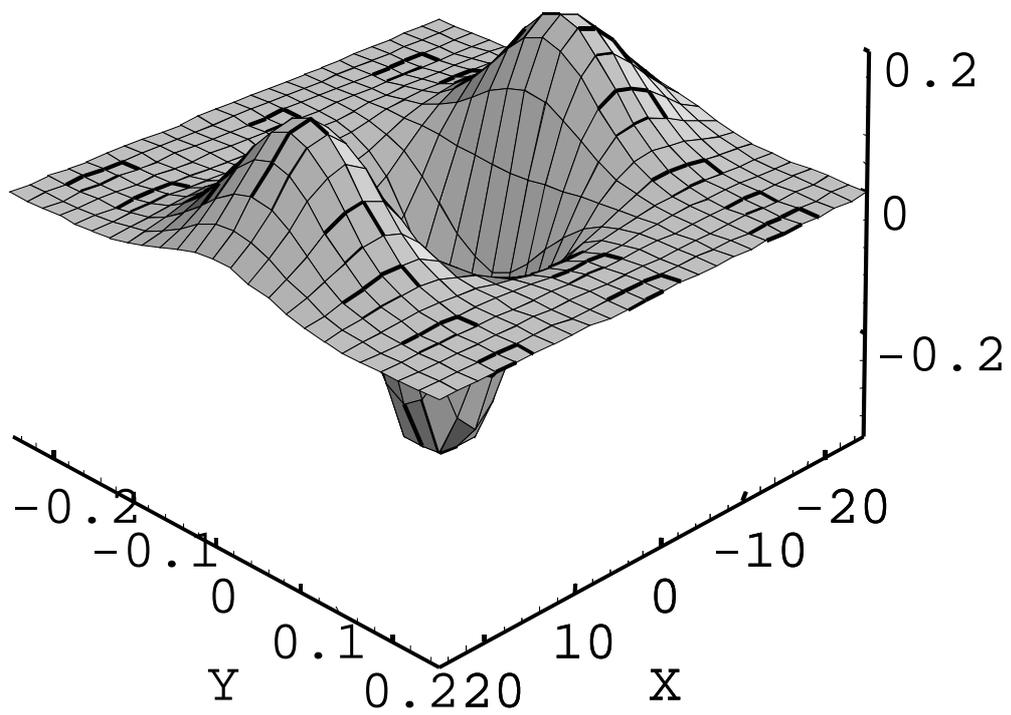
\end{document}